\documentclass[10pt, a4paper]{article}
	
\usepackage{amsmath, amsthm, amscd, amsfonts, amssymb, graphicx, color}
\usepackage[bookmarksnumbered, plainpages]{hyperref}

\newtheorem{thr}{Theorem}
\newtheorem{lem}{Lemma}
\newtheorem{cor}{Corollary}
\numberwithin{equation}{section}
\begin{document}
\centerline{\Large{\bf Efficient hedging in general Black-Scholes model}}
\centerline{}
\centerline{\textsuperscript{a}Kyong-Hui Kim, Myong-Guk Sin}
 \centerline{}
 \small \centerline{Faculty of Mathematics, \textbf{Kim Il Sung} University, D.P.R Korea}
\small \centerline{\textsuperscript{a}Corresponding author. e-mail address: kim.kyonghui@yahoo.com}
\centerline{}
\centerline{}
\begin{abstract}
An investor faced with a contingent claim may eliminate risk by perfect hedging, but as it is often quite expensive, he seeks partial hedging (quantile hedging or efficient hedging) that requires less capital and reduces the risk. Efficient hedging for European call option was considered in the standard Black-Scholes model with constant drift and volatility coefficients. In this paper we considered the efficient hedging for European call option in general Black-Scholes model $dX_t=X_t(m(t)dt+\sigma (t)dw(t))$ with time-varying drift and volatility coefficients and in fractional Black-Scholes model $dX_t=X_t(\sigma B_H(t)+mdt)$ with constant coefficients.
\end{abstract}

\hspace{1.2cm} time-varying coefficients, fractional Black-Scholes model  \\

%
%
\section{Introduction}

\quad In the standard Black-Scholes model with constant drift $m$ and  volatility $\sigma $, the underlying discounted price process is given by a geometric Brownian motion \cite{fol1}-\cite{gia}
\begin{equation}
dX_t=X_t(\sigma dw_t+mdt) \label{eq1.1}
\end{equation}
with initial value $X_0=x_0.$ A European call option $H=(X_T-K)^+$ can be hedged perfectly if an investor provides the initial capital
\begin{equation*}
U_0=E^*[H]=x_0\Phi (d_+)-K\Phi(d_-),
\end{equation*}
where
\begin{equation*}
 d_\pm (x_0,K)=\frac{\ln{x_0}-\ln{K}}{\sigma\sqrt{T}}\pm \frac{1}{2}\sigma\sqrt{T}.
\end{equation*}
In this case, an efficient hedging that minimizes the shortfall risk $E[\emph{l}((H-V_T)^+)]$ is equivalent essentially with the perfect hedging for modified claim $\tilde{H}=\tilde {\varphi }H.$
Thus \cite{fol1} makes clear that the efficient hedging is the problem for detecting $(V_0, \xi _s)$ that satisfies the equation
\begin{equation}
V_t=E^*[\tilde {\varphi }H\mid \Im _t ]=V_0+\int_0^{t}\xi _sdX_s, \quad \forall t\in [0,T], \label{eq1.2}
\end{equation}
and is a solution of the optimization problem 
\begin{equation}
E\left[\emph{l}((H-V_T)^+)\right]=E\left[\emph{l}((H-V_0-\int_0^{T}\xi _sdX_s)^+)\right]\Rightarrow min \label{eq1.3}
\end{equation}
under the constraint $V_0\leq \tilde {V_0}<U_0$.

In \cite{das, jac} pricing formulas for various options in Black-Scholes model 
\begin{equation}
dX_t=X_t(m(t)dt+\sigma (t)dw(t)) \label{eq1.4}
\end{equation}
with time-varying coefficients $m(t),\sigma (t)$ have been derived. The pricing currency options in a fractional Brownian motion with jumps and the option pricing for various fractional version of Black-Scholes model were considered in \cite{xia} and \cite{nec}-\cite{wan3}, respectively.  
In Ito-type fractional Black-Scholes model, underlying discounted price process satisfies the equation
\begin{equation}
dX_t=X_t(\sigma B_H(t)+mdt), \label{eq1.5}
\end{equation}
where  $B_H(t)$ is a fractional Brownian motion with Hurst parameter of $1/2<H<1$.

In this paper we consider the efficient hedging for European call option in two general Black-Scholes models \eqref{eq1.4} and \eqref{eq1.5} and extend the results for model \eqref{eq1.1} to models \eqref{eq1.4} and \eqref{eq1.5}. The rest of this paper is organized as follows. After giving preliminaries on efficient hedging and fractional Black-Scholes pricing formula in section 2, we present main results in section 3 and prove these results in section 4.

%
\section{Preliminaries}

%
\subsection{Efficient hedging}

Let the discounted price process of the underlying asset is described as a semi-martingale $X=(X_t)_{t\in [0,T]}$ on a probability space $(\Omega ,\mathbf{F}, P)$ with filtration $(\Im _t)_{t\in [0,T]}$. It is well known that in the complete market where the equivalent martingale measure is unique, $U_0=E^*[H]$ is the unique arbitrage-free price (perfect hedge price) of the contingent claim $H$. If the investor is unwilling or unable to put up the initial capital $U_0$, what is the best hedge the investor can achieve with a given smaller amount $\tilde {V_0}<U_0 $? In quantile hedging, they are looking for an admissible strategy $(V_0,\xi )$ which minimizes the probability of a shortfall  $P(V_T \leq H)$ under the constraint $V_0\leq \tilde {V_0}$ . In efficient hedging, they want to control not only the probability  $P(V_T \leq H)$  that some shortfall $(H-V_T )^+$occurs, but also the size of the shortfall, that is shortfall risk $E[\emph{l}((H-V_T)^+)]$. Their aim is to find an admissible strategy  $(V_0,\xi )$ which minimizes the shortfall risk while not using more than $\tilde {V_0}$. 
Thus in this paper we consider optimization problem \eqref{eq1.3}. This problem can be considered as choosing a function $\tilde {\varphi }\in \Phi $ such that
\begin{equation}
E\left[\emph{l}(1-\varphi )H\right]\Rightarrow min \label{eq2.1}
\end{equation}
under the constraint 
\begin{equation}
\underset{P^*\in P}{\text{sup}}E^*(\varphi H)<\tilde {V_0} \label{eq2.2}
\end{equation}
where $\Phi:=\{\varphi \mid \varphi :\Omega \rightarrow [0,1],\Im-\text{measurable}\}$. In case of complete market we can express $\tilde {\varphi }$ clearly in terms of given loss function $\emph{l}$ using Neyman-Pearson lemma. 

\begin{lem}\label{lem1} \textnormal{(\cite{fol1})}
The solution $\tilde {\varphi }$ of the optimization problem \eqref{eq2.1}, \eqref{eq2.2} is given by
\begin{equation*}
\tilde {\varphi }=1-\left(\frac{I(C\rho ^*)}{H}\wedge 1\right)
\end{equation*}
on $\{H>0\}$, where the constant $C$ is determined by the condition
\begin{equation*}
E^*[\tilde {\varphi }H]=\tilde {V_0}
\end{equation*}
and $\rho ^{*}=\frac{dP^*}{dP}$, $I=(l {'})^{-1}$.
\end{lem}

Let consider a loss function $\emph{l}(x)=\frac{X^p}{p} (p>1).$
 
\begin {lem}\label{lem2} \textnormal{(\cite{fol1})}
In case of\quad $\l(x)=\frac{X^p}{p},$ $\tilde {\varphi }_p=1-{\frac{C_p(\rho ^*)^{\frac{1}{p-1}}}{H}}\wedge 1$ and constant $C_p$ is determined by $ E^*[\tilde {\varphi }_pH]=\tilde {V_0}$.
\end{lem}
If we assume $\emph{l}(x)=x$ then this is a special case of loss functions that $I=(\emph{l} ')^{-1}$ doesn't exist.

\begin{lem}\label{lem3} \textnormal{(\cite{fol1})}
In case of\quad $\l(x)=x$, 
\begin{equation*}
\tilde {\varphi _1}=I\left\{\frac{dP}{dP^*}>\tilde {a}\right\} = I\left\{\rho^*<\frac{1}{\tilde {a}}\right\}
\end{equation*}
and constant $\tilde {a}$ is determined by
\begin{equation*}
\tilde {a}=\textnormal{inf}\left\{a \left\vert \int_{\left\{\frac{dP}{dP^*}>a\right\}} HdP^*\leq \tilde {V_0}\right. \right\}.
\end{equation*}
\end{lem} 

%

\subsection{Fractional Black-Scholes pricing formula}
Let the discounted price process of the underlying asset is described by Eq.\eqref{eq1.5}. Define 
\begin{equation*}
\theta =\frac{m}{\sigma }, ~ Z_t=\exp{\left(-\theta B_H(t)-\frac{\theta ^{2H}}{2}\right)}, ~ \frac{d\mu ^*}{d\mu }=Z_T, 
\end{equation*}
then $B_H^*(t)=B_H(t)+\theta t^{2H}$ is a fractional Brownian motion and $X_t$ is a quasi-martingale under the risk-neutral measure $\mu ^*$.The solution of the Eq.\eqref{eq1.5} is expressed as follows:
\begin{equation}
X_T=X_t\exp\left[\sigma (B_{H}^{*}(T)-B_{H}^{*}(t))-{\frac{1}{2}}\sigma ^{2}(T^{2H}-t^{2H})\right]. \label{eq2.3}
\end{equation}
We will denote by $\tilde {E}_t[\cdot ]$ the quasi-conditional expectation with respect to $\mu ^*$.

\begin{lem}\label{lem4} \textnormal{(\cite{nec})(fractional risk-neutral evaluation)}
In model \eqref{eq1.5}, the price at every $t\in [0,T]$ of a bounded $\Im _{T}^{H}-$ measurable claim $H\in L^{2}(\mu )$ is given by
\begin{equation}
V_t=\tilde {E}_t[H]. \label{eq2.4}
\end{equation}
\end{lem}

\begin{lem}\label{lem5} \textnormal{(\cite{nec})(fractional Black-Scholes formula)}
The price at every  $t\in [0,T]$ of European call option is given by
\begin{eqnarray*}
&& V_t=X_t\Phi (d_+)-K\Phi (d_-) \\
&& d_\pm =\frac{{\ln\left(\frac{X_t}{K}\right)}}{\sigma \sqrt{T^{2H}-t^{2H}}}\pm \frac{\sigma \sqrt{T^{2H}-t^{2H}}}{2}.
\end{eqnarray*}
\end{lem}

%

\section{Main results}
Let us consider a loss function $l(x)=\frac{1}{p}x^p$.
%
\begin{thr}\label{theo1}
Let us consider European call option $H=(X_T-K)^+$ in the general market model \eqref{eq1.4}. Then efficient hedging strategy  is as follows: 
\begin{eqnarray*}
 && V_t=F_{p}(t,X_t), \\
 && F_{p}(t,x)=x\Phi (d_+ (x,L))-K\Phi (d_-(x,L))-(L-K){\left(\frac{L}{x}\right)}^{\frac{\alpha _T}{p-1}} \\
 && \qquad \times \exp\left\{\frac{1}{2}\sigma _{T}^{2}{\frac{\alpha _T}{p-1}}\left({\frac{\alpha _T}{p-1}}+1\right)\right\}\Phi \left(d_-(X,L)-{\frac{\alpha_T\sigma _T}{p-1}}\right),\\
 && \xi _{p}(t,x)=\frac{\partial{F_{P}(t,x)}}{\partial{x}}=\Phi \left({\frac{\ln{x}-\ln{L}}{\sigma _T}}+\frac{1}{2}\sigma _{T}\right) +\frac{\alpha _T}{p-1}\frac{L^{\frac{\alpha _T}{p-1}}(L-K)}{x^{\frac{\alpha _T}{p-1}+1}}  \\
&& \qquad \times \exp \left\{ \frac{1}{2}\sigma _T^2 \frac{\alpha _T}{p-1} \left(\frac{\alpha _T}{p-1}+1\right)\right\} \Phi \left(\frac{\ln x-\ln L}{\sigma _T}-\sigma _T\left(\frac{\alpha _T}{p-1}+\frac{1}{2}\right)\right),
\end{eqnarray*} 
where
\begin{equation}
\sigma _T:=\left( \int_t^{T} \sigma ^2 (s) ds \right)^{\frac{1}{2}}, \quad \alpha _T:=\left( \frac{\int_t^{T} \theta ^2 (s) ds}{\int_t^{T} \sigma ^2 (s) ds} \right)^{\frac{1}{2}}, \quad \theta (t)=\frac{m(t)}{\sigma (t)}, \label{eq3.1}
\end{equation}
\begin{equation}
d_\pm (X, L) = \frac{\ln x - \ln K}{\sigma _T} \pm \frac{1}{2} \sigma _T. \label{eq3.2}
\end{equation}
\end{thr}

Note that the case of $l(x)=x$  is not a direct corollary of Theorem \ref{theo1}.
In this case the optimization strategy that minimizes  $E[(H-V_T)^+]$ is given as follows.

\begin{thr}\label{theo2}
In the case of\quad $l(x)=x$ efficient hedging strategy $(V, \xi )$ is as follows: 
\begin{equation}
V_t = F_1(t, X_t), D = \left( \frac{1}{\sigma _T} \ln \frac{K}{x}+\frac{1}{2}\sigma _T \right) \wedge \left(\frac{1}{\theta _T}\ln \tilde{a}+\frac{1}{2} \theta _T \right), \label{eq3.3}
\end{equation}
\begin{equation*}
F_1(t, x) = x \Phi (\sigma _T-D)-K \Phi (-D),
\end{equation*}
where the constant $\tilde{a}$ is determined by $\tilde{V}_0 = E^* [(X_T - K)I_A]$ with $A= \left[ \frac{dP}{dP^*} > \tilde{a} \right]$.
\end{thr}

\begin{thr}\label{theo3}
Let us consider European call option $H=(X_T-K)^+$ in the fractional Black-Scholes model \eqref{eq1.5}. Then efficient hedging strategy $(V, \xi )$ is given by replacing $\sigma _T, \alpha _T$  in \eqref{eq3.1}-\eqref{eq3.3} with
\begin{equation}
\sigma _T = \sigma \sqrt{T^{2H} - t^{2H}}, \quad \alpha _T = \frac{m}{\sigma ^2} \label{3.4}
\end{equation}
\end{thr}
\begin{cor}\label{cor1}
Let us consider European call option $H=(X_T-K)^+$ in the standard Black-Scholes model \eqref{eq1.1}. Then efficient hedging strategy $(V, \xi )$ is given by replacing $\sigma _T, \alpha _T$ in \eqref{eq3.1}-\eqref{eq3.3} with
\begin{equation}
\sigma _T = \sigma \sqrt{T - t}, \quad \alpha _T = \frac{m}{\sigma ^2} \label{eq3.5}
\end{equation}
\end{cor}
\textbf{Remark}. If we assume that $\sigma (t) \equiv \sigma, ~ m(t)\equiv m$ in model \eqref{eq1.4} or $H=1/2$ in model \eqref{eq1.5}, then these models are the same as model \eqref{eq1.1}.  Thus it follows that \eqref{eq3.5} holds from Theorem \ref{theo1}-Theorem \ref{theo3}. Corollary \ref{cor1} is no more than results of \cite{fol1}. We can see that Theorem \ref{theo1}-Theorem \ref{theo3} extend efficient hedging results of \cite{fol1} to general models \eqref{eq1.4} and \eqref{eq1.5}.

%
\section{Proofs of the theorems}

%
%
\subsection{Proof of Theorem \ref{theo1}} The solution of the stochastic differential equation \eqref{eq1.4} under equivalent martingale measure $P^*$ is as follows:
\begin{equation*} 
X_T = X_t \textrm{exp} \left\{ -\frac{1}{2} \int_t ^{T} \sigma ^2(s)ds+\int_t ^{T} \sigma (s)dw^*(s)\right\}
\end{equation*}
The unique equivalent martingale measure $P^*$ is given by
\begin{equation*} 
\frac{dP^*}{dP} = \rho ^* = \textrm{exp} \left\{ -\frac{1}{2} \int_0 ^{T} \theta ^2(s)ds-\int_0 ^{T} \theta (s)dw(s)\right\}
\end{equation*}
The process $w^*$ difined by $w^*(t) = w(t) + \int_0 ^{t} \theta(s)ds$ is a Brownian motion and $\left( X_t \right)$ is a martingale under measure $P^*$, where $\theta (s) = \frac{m(s)}{\sigma (s)}$. Let define
\begin{equation*} 
Z_t = \textrm{exp} \left\{ \frac{1}{2} \int_0 ^{t} \theta ^2(s)ds-\int_0 ^{t} \theta (s)dw^*(s)\right\},
\end{equation*}
then
\begin{equation*} 
\rho ^* = Z_T = Z_t \textrm{exp} \left\{ \frac{1}{2} \int_t ^{T} \theta ^2(s)ds-\int_t ^{T} \theta (s)dw^*(s)\right\}.
\end{equation*}
The distribution of $\int_t ^{T} \sigma (s)dw^*(s)$ is normal distribution with zero mean and variance $\int_t ^{T} \sigma ^2 (s)ds$, so we have
\begin{equation*}
\int_t ^{T} \sigma (s)dw^*(s) = \sqrt{\int_t ^{T} \sigma ^2 (s)ds} \cdot \tau
\end{equation*}
and similarly
\begin{equation*}
\int_t ^{T} \theta (s)dw^*(s) = \sqrt{\int_t ^{T} \theta ^2 (s)ds} \cdot \tau, 
\end{equation*}
where $\tau $ is standard normal distributed random variable independent with $\Im _t$. Using the definition of $\sigma_T$ and $\theta_T$, we obtain
\begin{equation*} 
X_T = X_t \textrm{exp} \left\{ \sigma _T \left( -\frac{\sigma_T}{2}+\tau \right) \right\}, \quad \rho ^* = Z_t exp \left\{ -\theta _T \left( -\frac{\theta_T}{2}+\tau \right) \right\}.
\end{equation*}
For European call option $H=(X_T-K)^+$, we have
\begin{equation*} 
\tilde{\varphi} _pH = (X_T-K)^+ - \left[ C^{\frac{1}{p-1}}(\rho ^*)^{\frac{1}{p-1}} \wedge (X_T-K)^+ \right].
\end{equation*}
Let define two functions
\begin{eqnarray*}
y_1 (x)&= C^{\frac{1}{p-1}} Z_t^{\frac{1}{p-1}} \textrm{exp} \left[ -\frac{\theta_T}{p-1} \left( -\frac{\theta_T}{2}+x \right) \right], \\
y_2 (x)&= \left( X_t \textrm{exp} \left[ \sigma_T \left( -\frac{\sigma_T}{2}+x \right) \right] -K \right) ^+.
\end{eqnarray*}

If we compare two functions $y_1$ and $y_2$, then there exists a unique solution $E$ such that $y_1=y_2$. If $x \geq E$ then $\tilde{\varphi} _pH = (X_T-K) - C^{\frac{1}{(p-1)}}(\rho^*)^{\frac{1}{(p-1)}}$ and if $x<E$ then $\tilde{\varphi} _pH = 0$. From the equation $y_1(E)=y_2(E)$,
\begin{equation*} 
C^{\frac{1}{p-1}} = \frac{X_t \textrm{exp} \left[ \sigma_T \left( -\sigma_T/2 +E \right) \right] -K}{Z_t^{1/(p-1)} \textrm{exp} \left[ -(\theta_T/(p-1))(-\theta_T/2 +E) \right]}.
\end{equation*}
Denote $A \left( \sigma _T, \theta_T, E, K, X_t \right) = C^{1/(p-1)}Z_t^{1/(p-1)}$, then
\begin{align*}
\tilde{\varphi} _pH &= \left[ (X_T-K) - C^{\frac{1}{p-1}}(\rho ^*)^\frac{1}{p-1} \right] I\{\tau\geq E\} \\
&= \left[ X_t \textrm{exp} \left\{ \sigma_T \left( -\frac{\sigma_T}{2}+\tau \right) \right\} -K-A \cdot \textrm{exp} \left\{ -\frac{\theta_T}{p-1} \left( -\frac{\theta_T}{2}+\tau \right) \right\} \right] \\
& \qquad \times I\{\tau \geq E\} \\
&= f_p (\tau)
\end{align*}
Let denote
\begin{align*} 
f_p(x) & = \left[ X_t \textrm{exp} \left\{ \sigma_T \left( -\frac{\sigma_T}{2}+x \right) \right\} -K-A \cdot \textrm{exp} \left\{ -\frac{\theta_T}{p-1} \left( -\frac{\theta_T}{2}+x \right) \right\} \right] \\
& \qquad \times I\{x \geq E\},
\end{align*}
then $\tilde{\varphi} _pH = f_p(\tau)$. Therefore
\begin{align*}
V_t &= E^* \left[ \tilde{\varphi} _pH | \Im _t \right] = E^* \left[ f_p(\tau) | \Im _t \right] \\
&= E^* \left[ \left[ X_t \textrm{exp} \left\{ \sigma_T \left( -\frac{\sigma_T}{2}+\tau \right) \right\} -K-A \cdot \textrm{exp} \left\{ -\frac{\theta_T}{p-1} \left( -\frac{\theta_T}{2}+\tau \right) \right\} \right] \right.\\
& \qquad \left. \times \vphantom{\left\{ -\frac{\theta_T}{p-1} \left( -\frac{\theta_T}{2}+\tau \right) \right\}} I\{\tau \geq E\} | \Im _t \right]  \\
&= F_p (t, X_t),
\end{align*}
where
\begin{align*}
F_p (t, x) &= \int_E^{+\infty } \left[ x \textrm{exp} \left\{ \sigma_T \left( -\frac{1}{2} \sigma _T+y \right) \right\} -K-A \cdot \textrm{exp} \left\{ -\frac{\theta_T}{p-1} \left( -\frac{1}{2}\theta _T +y \right) \right\} \right] \\
& \qquad \times \frac{1}{\sqrt{2\pi }} e^{- \frac{y^2}{2}}dy =: B_1 + B_2 + B_3
\end{align*}
and
\begin{align*} 
B_1 & = \frac{1}{\sqrt{2\pi }} \int _E^{+\infty } x\textrm{exp} \left\{ -\frac{\sigma_T^2}{2}+\sigma_T y-\frac{y^2}{2} \right\}dy \\
& = \frac{1}{\sqrt{2\pi }} x \int _{E-\sigma_T}^{+\infty } e^{-z^2/2} dz = x \Phi (\sigma _T-E), \\
B_2 &= -K \Phi (-E), \\
B_3 &= -A \cdot \frac{1}{\sqrt{2\pi }} \int _E^{+\infty } \textrm{exp} \left\{ \frac{\theta_T^2}{2(p-1)}-\frac{\theta_T}{p-1} y-\frac{1}{2} y^2 \right\}dy \\
&= -\frac{A}{\sqrt{2\pi }} \cdot \textrm{exp} \left\{ \frac{p \theta_T^2}{2(p-1)^2}\right\} \cdot \int _E^{+\infty } \textrm{exp} \left\{ -\frac{1}{2} \left( y+\frac{\theta_T}{p-1} \right)^2 \right\} dy \\
&= -A \cdot \textrm{exp} \left\{ \frac{1}{2} \cdot \frac{p\theta_T^2}{(p-1)^2} \right\} \cdot \Phi \left( -\frac{\theta_T}{p-1}-E \right).
\end{align*}
Let denote $L$ the value of $y_2$ at $x=E$, then $y_2(E)= X_t \textrm{exp} \left\{ \sigma_T \left( -\frac{\sigma_T}{2}+E \right) \right\} = L$, and thus $E=\frac{1}{\sigma_T} \textrm{ln} \left( \frac{L}{X_t} \right) +\frac{\sigma_T}{2}$.
Thus
\begin{align*}
F_p(t, x) &= x \cdot \Phi \left( \frac{\textrm{ln} x-\textrm{ln} L}{\sigma_T}+\frac{\sigma_T}{2} \right) -K \Phi \left( \frac{\textrm{ln} x-\textrm{ln} L}{\sigma_T}-\frac{\sigma_T}{2} \right) \\
&- A \cdot \textrm{exp} \left\{ \frac{1}{2} \frac{p\theta_T^2}{(p-1)^2} \right\} \cdot
\Phi \left( \frac{\textrm{ln} x-\textrm{ln} L}{\sigma_T}- \frac{\sigma_T}{2}-\frac{\theta_T}{p-1}\right).
\end{align*}
Now we calculate the third term of the above equation.
\begin{align*}
&A \cdot \textrm{exp} \left\{ \frac{1}{2} \frac{p\theta_T^2}{(p-1)^2} \right\} = (L-K) \textrm{exp} \left\{ \frac{p\theta_T^2}{2(p-1)^2}+\frac{\theta_T}{p-1} \left( -\frac{1}{2}\theta_T +E \right)\right\} \\
& \qquad =(L-K)\left(\frac{L}{x}\right)^{\frac{\theta_T}{(p-1)\sigma_T}} \textrm{exp} \left\{ \frac{p\theta_T^2}{2(p-1)^2}-\frac{\theta_T^2}{2(p-1)} + \frac{\theta_T \sigma_T}{2(p-1)} \right\} \\
& \qquad =(L-K)\left(\frac{L}{x}\right)^{\frac{1}{p-1} \cdot \frac{\theta_T}{\sigma_T}} \textrm{exp} \left\{ \frac{1}{2}\frac{\theta_T\sigma_T}{(p-1)} \left( \frac{\theta_T}{(p-1)\sigma_T}+1 \right) \right\}.
\end{align*}
Using above equation we have
\begin{align*}
F_p(t, x) &= x \Phi \left( \frac{\textrm{ln} x-\textrm{ln} L}{\sigma_T}+\frac{1}{2}\sigma_T \right) -K\Phi \left( \frac{\textrm{ln} x-\textrm{ln} L}{\sigma_T}-\frac{1}{2}\sigma_T \right) \\
&-(L-K)\left(\frac{L}{x}\right)^{\frac{1}{p-1} \cdot \frac{\theta_T}{\sigma_T}} \cdot \textrm{exp} \left\{ \frac{1}{2}\frac{\theta_T\sigma_T}{(p-1)} \left( \frac{\theta_T}{(p-1)\sigma_T}+1 \right) \right\} \\
& \times \Phi \left( \frac{\textrm{ln} x-\textrm{ln} L}{\sigma_T}-\frac{1}{2}\sigma_T -\frac{\theta_T}{(p-1)} \right).
\end{align*}
To obtain the results similar to \cite{fol1}, we use the definitions of $d_{\pm }(x, L), \alpha _T$, then
\begin{align*}
F_p(t, x) &= x \Phi (d_+(x, L))-K\Phi (d_-(x, L))-(L-K)\left(\frac{L}{x}\right)^{\frac{\alpha _T}{p-1}} \cdot \\
& \cdot \textrm{exp} \left\{ \frac{1}{2} \sigma_T^2 \frac{\alpha _T }{p-1} \left( \frac{\alpha _T }{p-1}+1 \right) \right\} \cdot \Phi \left( d_-(x, L)-\frac{\alpha_T\sigma_T}{p-1} \right).
\end{align*}
Let define
\begin{equation*} 
B:=\textrm{exp} \left\{ \frac{1}{2} \sigma_T^2 \frac{\alpha _T }{p-1} \left( \frac{\alpha _T }{p-1}+1 \right) \right\},
\end{equation*} 
then
\begin{align*}
\xi _p(t, x) &= \Phi (d_+(x, L)) + x \Phi' (d_+(x, L))-K\Phi' (d_-(x, L))+B \frac{\alpha _T}{p-1}L^{\frac{\alpha _T}{p-1}} \frac{(L-K)}{x^{\frac{\alpha _T}{p-1}+1}}  \\
& \quad \times \Phi \left( d_-(x, L)-\frac{\sigma_T \alpha _T}{p-1}\right) -B (L-K)\left(\frac{L}{x}\right)^{\frac{\alpha _T}{p-1}}\Phi' \left( d_-(x, L)-\frac{\sigma_T \alpha _T}{p-1}\right) \\
& = \Phi (d_+(x, L)) + x \Phi' (d_+(x, L))-L\Phi' (d_-(x, L))+(L-K)\Phi' (d_-(x, L)) \\
& \quad + B \frac{\alpha _T}{p-1}L^{\frac{\alpha _T}{p-1}} \frac{(L-K)}{x^{\frac{\alpha _T}{p-1}+1}}
 \cdot \Phi \left( d_-(x, L)-\frac{\sigma_T \alpha _T}{p-1}\right) \\
& \quad -B \cdot \frac{L^{\frac{\alpha _T}{p-1}}(L-K)}{x^{\frac{\alpha _T}{p-1}}} \Phi' \left( d_-(x, L)-\frac{\sigma_T \alpha _T}{p-1}\right).
\end{align*}
Let denote
\begin{align*}
& J_1 := x \Phi' (d_+(x, L))-L\Phi' (d_-(x, L)) \\
& J_2 := (L-K)\Phi' (d_-(x, L)) - B \cdot (L-K) \left( \frac{L}{x}\right)^{\frac{\alpha _T}{p-1}} \Phi' \left( d_-(x, L)-\frac{\sigma_T \alpha _T}{p-1}\right).
\end{align*}
Then $J_1=0$, because $\textrm{ln} \left( \frac{L}{x} \right) = -\frac{1}{2} \left[ d_+^2(x, L)-d_-^2(x, L) \right]$ from definition of $d_\pm (x, L)$ and
\begin{equation*} 
J_1 = \frac{1}{\sqrt{2\pi }} \left[ x\textrm{exp} \left( -\frac{1}{2}d_+^2(x, L) \right) -L\textrm{exp} \left( -\frac{1}{2}d_-^2(x, L) \right) \right].
\end{equation*}
Also we have
\begin{equation*} 
J_2 = \frac{(L-K)}{\sqrt{2\pi }} \left[ \textrm{exp} \left( -\frac{1}{2}d_-^2(x, L) \right) -B\left( \frac{L}{x}\right)^{\frac{\alpha _T}{p-1}}\textrm{exp} \left( -\frac{1}{2}\left( d_-(x, L)-\frac{\alpha _T \sigma _T}{p-1} \right)^2 \right) \right]\quad\quad\quad\quad\quad\quad\quad\quad.
\end{equation*}
Calculating the difference of indexes in same way as in the calculation of $J_1$ implies that
\begin{equation*} 
-\frac{1}{2} \left\{ d_-^2(x, L)-\left( d_-(x, L)-\frac{\alpha _T \sigma _T}{p-1} \right)^2 \right\} = -\frac{1}{2} \left\{ 2 \cdot \frac{\alpha _T \sigma _T}{p-1} d_-(x, L)-\frac{\alpha _T^2 \sigma _T^2}{(p-1)^2} \right\}
\end{equation*}
\begin{equation*} 
=\frac{1}{2} \left(\frac{\alpha _T \sigma _T}{p-1} \right)^2 - \frac{\alpha _T \sigma _T}{p-1} \left[ \frac{\textrm{ln} x-\textrm{ln} L}{\sigma_T}-\frac{1}{2}\sigma _T \right] = \textrm{ln} \left( B \cdot \left( \frac{L}{x}\right)^{\frac{\alpha _T}{p-1}} \right).
\end{equation*}
Therefore $J_2=0$, which proves Theorem \ref{theo1}. 

%
%
\subsection{Proof of Theorem 2}

From Lemma \ref{lem3} we easily see that
\begin{align*}
V_t &= F_1(t, X_t) = E^* \left[ \tilde{\varphi} _1H | \Im _t \right] = E^* \left[ HI_A | \Im _t \right] \\
&= E^* \left\{ \left[ X_t \textrm{exp} \left\{ \sigma_T \left( -\frac{\sigma_T}{2}+\tau \right) \right\} -k\right] I \left\{\tau \geq \left( \frac{1}{\sigma _T} \textrm{ln} \frac{k}{X_t} + \frac{1}{2} \sigma _T \right) \right. \right.\\
& \qquad \qquad \qquad \qquad \left. \left. \left. \wedge \left( -\frac{1}{\theta _T} \textrm{ln} \frac{1}{\tilde{a}} + \frac{1}{2} \theta _T \right) \right\} \right\vert \Im _t \right\} \\
&= X_t \Phi (\sigma _T-D)-k\Phi (-D),
\end{align*}
where
\begin{align*}
& D = \left( \frac{1}{\sigma _T} \textrm{ln} \frac{k}{X_t} + \frac{1}{2} \sigma _T \right) \wedge \left( \frac{1}{\theta _T} \textrm{ln} \tilde{a} + \frac{1}{2} \theta _T \right), \\
& \sigma _T-D = \left( \frac{1}{\sigma _T} \textrm{ln} \frac{X_t}{k} + \frac{1}{2} \sigma _T \right) \vee \left( -\frac{1}{\theta _T} \textrm{ln} \tilde{a} + \sigma _T -\frac{1}{2} \theta _T \right).
\end{align*}
Thus we obtain the result of Theorem \ref{theo2}. 

%
%
\subsection{Proof of Theorem \ref{theo3}}

The solution $X_T$ of Eq.\eqref{eq1.5} is given by \eqref{eq2.3}. As in standard Black-Scholes model, efficient hedging is equivalent essentially with perfect hedging for modified claim $\tilde{H} = \tilde{\varphi}H$. Thus efficient hedging price is $V_t = \tilde{E}_t [\tilde{\varphi}H]$ from Lemma \ref{lem5} in Ito-type fractional Black-Scholes model. For European call option $H=(X_t - K)^+$ we can prove Theorem \ref{theo3} by following similar method to proof of Theorem \ref{theo1}, Theorem \ref{theo2} and Lemma \ref{lem5}. 

%
%
\section{Conclusion}

In this paper we derived efficient hedging formulae for European call option in two general Black-Scholes models, so extended the results in the standard Black-Scholes model to general cases.

\end{document}